\documentclass[epj]{svjour}
\usepackage{graphics}
\usepackage{amsmath}
\usepackage{amsopn}
\usepackage{amssymb}
\usepackage{epsfig}
\usepackage{latexsym}

\newcommand{\bra}[1]{\langle #1|}
\newcommand{\ket}[1]{|#1\rangle}

\begin{document}

\title{Cotunneling through a quantum dot coupled to ferromagnetic
leads with noncollinear magnetizations}

\author{Ireneusz Weymann\inst{1} \and J\'ozef Barna\'s\inst{1,2}}

%
%
\institute{Department of Physics, Adam Mickiewicz University,
61-614 Pozna\'n, Poland \and Institute of Molecular Physics,
Polish Academy of Sciences, 60-179 Pozna\'n, Poland}

\date{Received: date / Revised version: date}

\abstract{ Spin-dependent electronic transport through a quantum
dot has been analyzed theoretically in the cotunneling regime by
means of the second-order perturbation theory. The system is
described by the impurity Anderson Hamiltonian with arbitrary
Coulomb correlation parameter $U$. It is assumed that the dot
level is intrinsically spin-split due to an effective molecular
field exerted by a magnetic substrate. The dot is coupled to two
ferromagnetic leads whose magnetic moments are noncollinear. The
angular dependence of electric current, tunnel magnetoresistance,
and differential conductance are presented and discussed. The
evolution of a cotunneling gap with the angle between magnetic
moments and with the splitting of the dot level is also
demonstrated.
} 

\maketitle

\section{Introduction}

The vast step forward in fabrication of tunnel junctions has
enabled the implementation of extremely small metallic grains and
semiconductor quantum dots coupled through tunnel barriers to
external reservoirs \cite{kouwenhoven97,kouwenhoven98}. Currently,
transport through such ultra-small devices is being extensively
studied because of many future application possibilities and, more
importantly, because of beautiful physics that emerges in these
nanoscale systems. In nanostructures not only the manipulation of
a single electron charge is possible,  but -- when coupled to
ferromagnetic leads -- also the manipulation of a single electron
spin. This is why those systems are considered to play an
important role in spintronic devices.

An interesting feature of electronic transport through nanoscale
systems coupled to ferromagnetic leads is the tunnel
magnetoresistance (TMR). The TMR effect, although discovered
already three decades ago in planar junctions \cite{julliere75},
is still of current interest. It consists in a change of the
system conductance when relative orientation of the magnetic
moments of external leads switches from antiparallel to parallel
alignment. In a general case, magnetic moments of the two
electrodes can form an arbitrary angle (noncollinear
configuration). Qualitatively different features of electron
transport through nanosystems appear due to discrete charging
effects, leading to Coulomb blockade and Coulomb oscillations of
electric current. The interplay of charge and spin effects gives
rise to new interesting phenomena, like for instance TMR
oscillations with applied bias and gate voltages
\cite{barnas98,Brataas99,majumdar98,weymann03,yakushiji02,ernult04}.

Theoretical considerations of electron tunneling through quantum
dots attached to ferromagnetic leads have already been reported in
a number of papers. Most of them, however, are restricted to
spin-dependent transport in systems with collinear alignment of
the electrodes' magnetizations. Basically, all transport regimes
in such a geometry have already been addressed, including
sequential (first order) tunneling \cite{bulka00,rudzinski01},
cotunneling (second order) \cite{weymann04}, resonant tunneling
\cite{swirkowicz02,utsumi05}, and Kondo regimes
\cite{martinek03,lopez03}. Spin polarized transport through
quantum dots coupled to ferromagnetic electrodes with noncollinear
magnetic moments is still not fully explored, although it has
already been considered in a couple of papers
\cite{sergueev02,konig03,braun04,rudzinski04,flensberg04,fransson05}.

In this paper we address the problem of second-order (cotunneling)
spin-dependent transport through quantum dots coupled to
ferromagnetic leads with arbitrary configuration of the in-plane
magnetic moments of external electrodes. The considerations are
limited to single-level quantum dots. Moreover, the level of the
corresponding isolated dot is assumed to be spin-split due to
coupling between the dot and a magnetic substrate on which the dot
is deposited. The corresponding splitting is assumed to be larger
than the level width $\Gamma$ due to coupling of the dot to
external leads.

The paper is organized as follows. The model and method are
described in section II. Transport through an empty quantum dot in
the cotunneling regime is described in section III, where also the
relevant numerical results are presented and discussed. Section IV
covers the problem of cotunneling through a singly occupied dot.
Final conclusions are given in section V.

\section{Description of the model and method}

The system considered in this paper consists of a single-level
quantum dot coupled through tunnel barriers to two external
ferromagnetic leads, whose magnetizations are oriented arbitrarily
in the plane of the structure. The dot is assumed to be deposited
on a ferromagnetic substrate which strongly interacts with the dot
and leads to spin-splitting of the dot level. The splitting is
assumed to be larger than the level splitting due to exchange
interaction between the dot and electrodes. The exchange
interaction results from tunneling processes and is of the first
order in the coupling parameter $\Gamma$ \cite{konig03,braun04}.
When neglecting the exchange coupling between the dot and leads,
one may assume that the level splitting is constant, i.e.,
independent of applied voltage.

\begin{figure}[t]
\begin{center}
\includegraphics[width=0.7\columnwidth]{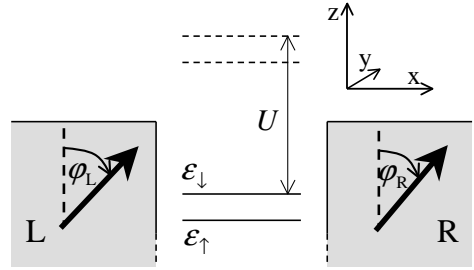}
  \caption{\label{Fig:1}The schematic of a magnetic quantum dot
  coupled to ferromagnetic leads. Classical spin of the lead
  $\alpha$ ($\alpha ={\rm L,R}$) can form an arbitrary angle
  $\varphi_\alpha$ with the dot spin quantization axis (axis $z$), as
  indicated.}
\end{center}
\end{figure}

Coupling of the dot to ferromagnetic substrate is described by an
effective molecular field ${\bf B}_s$, lying in the plane of the
structure. Thus, the magnetic moments of the electrodes and the
molecular field are all in a common plane. The effective molecular
field determines the global quantization axis for electron spin on
the dot (axis $z$ antiparallel to the molecular field). In turn,
the local quantization axis in the leads is determined by the
corresponding local magnetization direction, and is parallel to
the net spin of the lead (thus, being antiparallel to the local
magnetic moment). Furthermore, we assume that the net spin of the
left (right) lead forms an angle $\varphi_{\rm L}$ $(\varphi_{\rm
R})$ with the global quantization axis as shown in Fig.
\ref{Fig:1}. In order to distinguish between different
quantization directions, the majority (minority) spins in the
local reference systems are labelled with $\sigma=+(-)$, while
spin projection on the global quantization axis is denoted as
$\sigma=\uparrow (\downarrow)$ for spin-up (spin-down) electrons.

Hamiltonian of the system  has the general form $H=H_{\rm
L}+H_{\rm R}+H_{\rm D}+H_{\rm T}$, where $H_{\rm L}$ and $H_{\rm
R}$ describe the left and right leads as reservoirs of
noninteracting quasi-particles, $H_{\rm D}$ is the dot
Hamiltonian, and tunneling processes between the electrodes and
dot are included in $H_{\rm T}$. The lead Hamiltonians are
diagonal in the respective local coordinate systems,
$H_{\alpha}=\sum_{\sigma=+,-}\sum_{k\in\alpha}\varepsilon_{\alpha
k\sigma} a_{\alpha k\sigma}^{\dagger}a_{\alpha k\sigma}$ (for
$\alpha={\rm L,R}$), with $\varepsilon_{\alpha k\sigma}$ being the
energy of a single electron with wavevector $k$ and spin $\sigma$
in the lead $\alpha$, whereas $a_{\alpha k\sigma}^{\dagger}$ and
$a_{\alpha k\sigma}$ denote the corresponding creation and
annihilation operators.

The dot is described by the Anderson Hamiltonian, which in the
global reference frame can be expressed as $H_{\rm D}
=\sum_{\sigma= \uparrow,\downarrow} \varepsilon_{\sigma}
d_{\sigma}^{\dagger} d_{\sigma}+Un_{\uparrow}n_{\downarrow}$,
where $\varepsilon_{\sigma}$ is the energy of an electron with
spin $\sigma$,  $\varepsilon_{\sigma} =  \varepsilon \mp
g\mu_BB_s$, and $d_{\sigma}^{\dagger}$ $(d_{\sigma})$ creates
(annihilates) a spin-$\sigma$ electron. Here, $g$ is the Lande
factor and  $\varepsilon$ is the dot level energy in the absence
of magnetic electrodes and molecular field due to the substrate.
The second term of the dot Hamiltonian describes Coulomb
interaction of two electrons of opposite spins residing on the
dot, with $U$ denoting the corresponding correlation energy. The
tunnel Hamiltonian takes the form
\begin{eqnarray}
  H_{\rm T}=\sum_{\alpha}\sum_{k\in\alpha} [( T_{\alpha k+}a_{\alpha
  k+}^{\dagger} \cos\frac{\varphi_{\alpha}}{2} - T_{\alpha
  k-}a_{\alpha
  k-}^{\dagger} \sin\frac{\varphi_{\alpha}}{2} )d_{\uparrow}\nonumber\\
  + ( T_{\alpha k+}a_{\alpha k+}^{\dagger}
  \sin\frac{\varphi_{\alpha}}{2} + T_{\alpha k-}a_{\alpha
  k-}^{\dagger} \cos\frac{\varphi_{\alpha}}{2} )d_{\downarrow} +
  {\rm h.c.} ],
\end{eqnarray}
with $T_{\alpha k\sigma}$ denoting the tunnel matrix elements
between the dot states and majority $(\sigma=+)$ or minority
$(\sigma=-)$ electron states in the lead $\alpha$ when
$\varphi_\alpha =0$.

Due to the coupling between dot and leads, the dot level acquires
a finite width. When the magnetic moment of lead $\alpha$ and the
molecular field acting on the dot are parallel, the corresponding
contribution $\Gamma^\sigma_\alpha$ to the dot level width may be
written as $\Gamma_{\alpha}^{\sigma}=2\pi |T_{\alpha
\sigma}|^2\rho_{\alpha \sigma}$, where $\rho_{\alpha \sigma}$ is
the spin-dependent density of states for the majority $(\sigma=+)$
and minority $(\sigma=-)$ electrons in the lead $\alpha$. The
parameters $\Gamma_{\alpha}^{\sigma}$ will be used in the
following to parameterize strength of the coupling between the dot
and lead $\alpha$. It is convenient to express the coupling
parameters in terms of spin polarization defined as $P_{\alpha} =
(\Gamma_{\alpha}^{+}-\Gamma_{\alpha}^{-})/
(\Gamma_{\alpha}^{+}+\Gamma_{\alpha}^{-})$. Thus, the coupling
strength can be written as
$\Gamma_{\alpha}^{\pm}=\Gamma_{\alpha}(1\pm P_{\alpha})$, where
$\Gamma_{\alpha}=(\Gamma_{\alpha}^{+}+\Gamma_{\alpha}^{-})/2$. In
our considerations we assume $\Gamma_{\rm L}=\Gamma_{\rm R}=
\Gamma/2$.

In the case of ferromagnetic leads, the coupling of the spin-up
dot level is different from the coupling of the spin-down level,
which is due to different densities of states for spin-majority
and spin-minority electron bands in the leads. This may result in
the splitting of the dot level \cite{weymann04,fransson02}. Here,
we assume that the dot level splitting, $\Delta =\varepsilon
_\downarrow -\varepsilon_\uparrow$, due to the molecular field is
larger than the coupling parameters, $\Delta\gg\Gamma^\pm_\alpha$.
In other words, we assume that the exchange interaction between
the dot and magnetic leads is much smaller than the Zeeman energy
due to the molecular field and can be neglected. An electron
residing on the dot has then either spin up or down. Thus, the
corresponding density matrix in the global quantization system is
diagonal in the spin space \cite{remark}. In that case, only the
second-order processes (with respect to the tunneling Hamiltonian)
have to be taken into account in the Coulomb blockade regime. When
the above condition is not fulfilled, the exchange interaction
cannot be ignored and therefore the first-order processes, which
are responsible for the exchange coupling (and do not contribute
to charge transport), have to be considered.

In the Coulomb blockade regime, the energy needed to put an
electron on the dot surpasses the energy provided by the transport
voltage and the sequential tunneling is exponentially suppressed.
The dot is then in a well defined charge state, and quantum charge
fluctuations are suppressed. Although the energy conservation
prohibits the first-order tunneling transitions, the current can
still be mediated by higher-order tunneling processes involving
correlated tunneling of two (cotunneling) or more electrons {\it
via} intermediate virtual states \cite{averin90}.

The rate of electron cotunneling from a spin-majority state in the
lead $\alpha$ to a spin-majority state in the lead $\beta$ can be
determined using the second-order perturbation theory
\cite{averin90,kang97}, and is given by
\begin{equation}\label{Eq:cotunnelingrate}
  \gamma_{\alpha\beta}^{+\Rightarrow +}=\frac{2\pi}{\hbar}
  \left|\sum_{v}\frac{\bra{\Phi_\alpha^+}H_{\rm T}\ket{\Phi_v}
  \bra{\Phi_v} H_{\rm T} \ket{\Phi_\beta^+}}
  {\varepsilon_i-\varepsilon_{v}}\right|^2\delta
  (\varepsilon_i-\varepsilon_f),
\end{equation}
where $\ket{\Phi_\alpha^+}$ and $\ket{\Phi_\beta^+}$ are the
initial and final states of the system, whereas  $\ket{\Phi_v}$ is
a virtual state. The corresponding energies are denoted as
$\varepsilon_i$,  $\varepsilon_f$, and $\varepsilon_{v}$.

One can distinguish cotunneling processes that change the magnetic
(and consequently also energetic) state of the dot, and processes
affecting neither magnetization nor energy of the dot. The former
(latter) processes will be referred to as inelastic (elastic)
ones. The elastic cotunneling processes are fully coherent
\cite{konig01} and do not change the dot occupation probabilities.
Contrarily, inelastic cotunneling influences the occupation
numbers of the dot, and can take place only when the dot is
occupied by a single electron -- either spin-up or spin-down.
Furthermore, one can also distinguish between single-barrier and
double-barrier cotunneling. Only the latter processes contribute
directly to the current. However, the inelastic single-barrier
cotunneling processes can change the occupation probabilities, and
consequently can also influence the electric current flowing
through the system. In particular, inelastic single-barrier
cotunneling processes which reverse spin of the dot play a
significant role. This is because they can open system for more
efficient tunneling processes, when the system is blocked by an
electron of a given spin orientation residing on the dot
\cite{weymann04}.

In the following we will consider two different situations. The
first one corresponds to an empty dot (due to particle-hole
symmetry the results can be adapted to the case of doubly occupied
dot). The second situation is the case of a singly occupied dot.

\section{Cotunneling through an empty dot}

When the dot level is far above the Fermi energy of the leads,
$\varepsilon_\sigma \gg k_{\rm B}T, \Gamma, \vert eV\vert$, there
are no electrons on the quantum dot and electric current can flow
only due to elastic cotunneling processes. Because of the
particle-hole symmetry, a similar analysis can be directly
performed for a doubly occupied dot, when $\varepsilon_\sigma + U
\ll 0$ and $\vert \varepsilon_\sigma +U \vert \gg k_{\rm B}T,
\Gamma, \vert eV\vert$. As in the case of an empty dot, the
current can then flow only due to elastic cotunneling.

\subsection{Theoretical description}

Electric current $I$ flowing from the left to right lead is given
by
\begin{eqnarray}
  I&=& -e \sum_{\sigma,\sigma^\prime=+,-}\left(\gamma_{\rm
  L R,0}^{\sigma \Rightarrow \sigma^\prime}-\gamma _{\rm R
  L,0}^{\sigma \Rightarrow \sigma^\prime }\right),
\end{eqnarray}
with $\gamma_{\rm LR,\chi }^{\sigma \Rightarrow \sigma^\prime}$
being the elastic cotunneling rate for transition from the left to
right leads, when the dot is in the state $\ket{\Phi_\chi}$ [$\chi
=0$ in Eq.(3)] and when a majority $(\sigma=+)$ or minority
$(\sigma=-)$ electron of the left lead tunnels to majority
$(\sigma^{\prime}=+)$ or minority $(\sigma^{\prime}=-)$ electron
band in the right lead. Similarly, $\gamma _{\rm R L,0}^{\sigma
\Rightarrow \sigma^\prime }$ is the elastic cotunneling rate for
transition from the right to left electrodes. In Eq. (3) $-e$
denotes the electron charge $(e>0)$.

The transition rate for electrons tunneling from the majority spin
band in the left lead to the majority spin band in the right lead
is given by the formula
\begin{eqnarray}
  \lefteqn{ \gamma_{\rm LR,0}^{+ \Rightarrow +} = \frac{2 \pi}{\hbar} \iint
  d\varepsilon_{\rm L} d\varepsilon_{\rm R}
  \rho_{\rm L+}\rho_{\rm R+}} \nonumber\\
  &&\times |T_{\rm L+}|^2|T_{\rm R+}|^2
  f(\varepsilon_{\rm L})[1-f(\varepsilon_{\rm R})] \nonumber\\
  &&\times\left[ \frac{\cos{(\varphi_{\rm L}/2)}\cos{(\varphi_{\rm
  R}/2)}} {\varepsilon_{\rm L}+\mu_{\rm L}-\varepsilon_{\uparrow}}+
  \frac{\sin{(\varphi_{\rm L}/2)}\sin{(\varphi_{\rm R}/2)}}
  {\varepsilon_{\rm L}+\mu_{\rm L}-\varepsilon_{\downarrow}}
  \right]^2 \nonumber\\
  &&\times \delta(\varepsilon_{\rm L}+\mu_{\rm L}-\varepsilon_{\rm
  R}- \mu_{\rm R}) \,,
\end{eqnarray}
with $\mu_{\rm L}$ ($\mu_{\rm R}$) denoting the electrochemical
potential of the left (right) lead and $f(\varepsilon)$ being the
Fermi-Dirac distribution function,
$f(\varepsilon)=1/[\exp(\varepsilon/k_{\rm B}T)+1]$. We assume
$\mu_{\rm L}=-eV/2$ and $\mu_{\rm R}=eV/2$, and the energy is
measured from the Fermi level of the leads in equilibrium
situation ($V=0$). The integrals in Eq.(4) can be calculated quite
easily using the contour integration method, as described in the
Appendix. Following this procedure one finds
\begin{eqnarray}
    \gamma_{\rm LR,0}^{+\Rightarrow +}&=&\frac{\Gamma_{\rm L}^{+}
    \Gamma_{\rm R}^{+}}{h}f_{\rm B}(\mu_{\rm R}-\mu_{\rm L})\left\{
    \cos^2\frac{\varphi_{\rm L}}{2} \cos^2\frac{\varphi_{\rm R}}{2}
    A_2(\varepsilon_{\uparrow}) \right. \nonumber\\
    && + \sin^2\frac{\varphi_{\rm L}}{2}
    \sin^2\frac{\varphi_{\rm R}}{2} A_2(\varepsilon_{\downarrow})
    \nonumber\\
    &&\left.+ \frac{\sin\varphi_{\rm L} \sin\varphi_{\rm
    R}}{2(\varepsilon_{\uparrow}-\varepsilon_{\downarrow})}
    \left[A_1(\varepsilon_{\uparrow})-A_1(\varepsilon_{\downarrow})
    \right] \right\} \,,
\end{eqnarray}
where $f_{\rm B}(\varepsilon)$ is the Bose function, $f_{\rm
B}(\varepsilon)=1/[\exp(\varepsilon/k_{\rm B}T)-1]$, and
$A_n(\varepsilon_{\sigma})=B_n(\varepsilon_{\sigma}-\mu_{\rm R})-
B_n(\varepsilon_{\sigma}-\mu_{\rm L})$, with $B_n(x)$ defined as
\begin{equation}
  B_{n}(x)={\rm Re}\frac{{\rm d}^{(n-1)}}{{\rm d}^{(n-1)}x}
  \Psi\left(\frac{1}{2}+\frac{{\rm i}x}{2\pi k_{\rm B} T} \right).
\end{equation}
Here $\Psi(z)$ is the digamma function. Similar formulae can also
be derived for the other transition rates.

Having found all the cotunneling rates, one can calculate the
electric current,
\begin{eqnarray}
  I&=&\frac{e\Gamma^2}{4h} \bigg\{
  (1-P_{\rm L}\cos\varphi_{\rm L})
  (1-P_{\rm R}\cos\varphi_{\rm R})
  A_2(\varepsilon_\downarrow)\nonumber\\
  &&+(1+P_{\rm L}\cos\varphi_{\rm L})
  (1+P_{\rm R}\cos\varphi_{\rm R})
  A_2(\varepsilon_\uparrow)\nonumber\\
  &&+\frac{2P_{\rm L}P_{\rm R}\sin\varphi_{\rm L}\sin\varphi_{\rm
  R}}{\varepsilon_\downarrow-\varepsilon_\uparrow}
  \left[A_1(\varepsilon_\downarrow)-A_1(\varepsilon_\uparrow)\right]\bigg\}.
\end{eqnarray}

In the following we will present numerical results on the electric
current and the associated magnetoresistance. The TMR effect is
described quantitatively by the ratio
\begin{equation}
  \text{TMR}=\frac{I_{\rm P}-I(\varphi_{\rm L},\varphi_{\rm R})}
  {I(\varphi_{\rm L},\varphi_{\rm R})},
\end{equation}
where $I(\varphi_{\rm L},\varphi_{\rm R})$ is the current flowing
in the noncollinear configuration described by the angles
$\varphi_{\rm L}$ and $\varphi_{\rm R}$, whereas $I_{\rm P}$ is
the current flowing in the parallel configuration corresponding to
$\varphi_{\rm L}= \varphi_{\rm R}=0$.

\subsection{Numerical results}

\begin{figure}[t]
\begin{center}
\includegraphics[width=1\columnwidth]{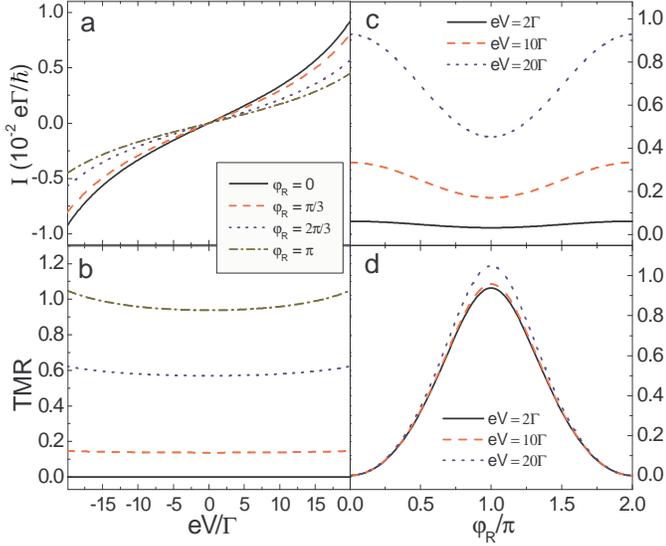}
  \caption{\label{Fig:2} (Color online) The cotunneling current
  (a,c) and the TMR effect (b,d) as a function of the bias voltage
  (left column) and the angle $\varphi_{\rm R}$ (right column). The
  parameters assumed for calculations are: $k_{\rm B}T=\Gamma$,
  $\varepsilon_{\uparrow}=18\Gamma$,
  $\varepsilon_{\downarrow}=22\Gamma$, $P_{\rm
  L}=P_{\rm R}=0.5$, $\varphi_{\rm L}=0$.}
\end{center}
\end{figure}

The formula (7) for electric current corresponds to the situation,
where the magnetic moments of the leads and the effective
molecular field acting on the dot are oriented arbitrarily in the
plane of the structure. Further numerical analysis will be
restricted, however, to the following two situations: (i) the
magnetic moment of the left lead is parallel to the molecular
field acting on the dot ($\varphi_{\rm L}=0$), while the magnetic
moment of the right lead can have an arbitrary orientation, and
(ii) the magnetic moments of both leads can rotate symmetrically
in the opposite directions, $\varphi_{\rm R}=-\varphi_{\rm L}$.
The case (i) corresponds to the situation when magnetic moment of
one lead and magnetic moment of the layer producing the molecular
field acting on the dot are fixed along the same direction. This
can be achieved for instance by a common antiferromagnetic
underlayer with strong exchange anisotropy at the
antiferromagnet/ferromagnet interface. The situation (ii), in
turn, corresponds to the case when both leads are equivalent and
their magnetic moments can be rotated simultaneously by an
external magnetic field. Now, we will analyze both situations in
more detail and we begin with the case (i).

\subsubsection{Case (i): $\varphi_{\rm L}=0$}

When $\varphi_{\rm L}=0$, the electric current is given by the
formula
\begin{eqnarray}
  I&=&\frac{e\Gamma^2}{4h} \left[(1-P_{\rm L})
  (1-P_{\rm R}\cos\varphi_{\rm R})
  A_2(\varepsilon_\downarrow)\nonumber\right.\\
  &&\left.+(1+P_{\rm L})(1+P_{\rm R}\cos\varphi_{\rm R})
  A_2(\varepsilon_\uparrow)\right],
\end{eqnarray}
which follows directly from Eq. (7). In Figure \ref{Fig:2} we show
the current (a,c) flowing through the system and the corresponding
TMR (b,d) as a function of the bias voltage for several values of
angle $\varphi_{\rm R}$, and as a function of the angle
$\varphi_{\rm R}$ for several values of the bias voltage. The
current decreases and TMR increases as the angle $\varphi_{\rm R}$
varies from $\varphi_{\rm R}=0$ to $\varphi_{\rm R}=\pi$, which
corresponds to the transition from parallel to antiparallel
magnetic configurations. Parts (c) and (d) show explicitly this
angular dependence. Both the current and TMR vary monotonously
with $\varphi_{\rm R}$ (for $0\le\varphi_{\rm R}\le\pi$) and
electric current reaches minimum, while TMR maximum, at
$\varphi_{\rm R}=\pi$, i.e., in the antiparallel configuration.
Such a behavior is typical of normal spin valves and results from
spin asymmetry in tunneling processes. It is also worth noting
that TMR is only weakly dependent on the bias voltage [see Fig.
2(b)].

Assuming the same spin polarization of the leads, $P_{\rm
L}=P_{\rm R}=P$, one finds the following explicit formula for TMR
in the zero bias and zero temperature limits:
\begin{equation}
  {\rm TMR}=\frac{(1-\cos\varphi_R)P\left[
    (1+P)\varepsilon_\downarrow^2-(1-P)\varepsilon_\uparrow^2
    \right]}
  {(1+P)(1+P\cos\varphi_R)\varepsilon_\downarrow^2+
  (1-P)(1-P\cos\varphi_R)\varepsilon_\uparrow^2}.
\end{equation}
The above formula describes the angular variation of TMR and shows
explicitly that TMR reaches maximum for $\varphi=\pi$. This
maximum value is given by the expression
\begin{equation}
{\rm TMR}^{\rm (max)}=\frac{2P}{1-P^2}
\left[P+\frac{\varepsilon_\downarrow^2-\varepsilon_\uparrow^2}
{\varepsilon_\uparrow^2+\varepsilon_\downarrow^2}\right].
\end{equation}
The first term in the bracket of the above equation gives the
Julliere's value of TMR, whereas the second term describes
enhancement of the tunnel magnetoresistance  due to level
splitting. Such an enhancement of TMR  may be of some interest
from the application point of view.

\subsubsection{Case (ii): $\varphi_{\rm R}=-\varphi_{\rm L}$}

\begin{figure}[t]
\begin{center}
  \includegraphics[width=1\columnwidth]{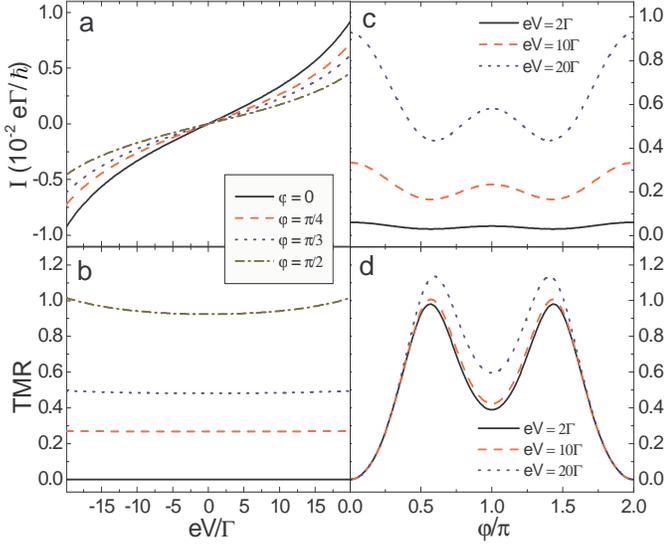}
  \caption{\label{Fig:3} (Color online) The cotunneling current
  (a,c) and the TMR effect (b,d) as a function of the bias voltage
  (left column) and as a function of $\varphi=\varphi_{\rm R}=-\varphi_{\rm L}$
  (right column). The other parameters are the same as in Fig. 2.}
\end{center}
\end{figure}

Consider now the situation (ii), when $\varphi_{\rm
R}=-\varphi_{\rm L} \equiv \varphi$. The magnetic moment of the
left lead rotates now together with the magnetic moment of the
right lead, but in the opposite direction. The electric current is
then given by Eq. (7) with $\varphi_{\rm R}= \varphi$ and
$\varphi_{\rm L} =- \varphi$. The corresponding bias and angular
dependence of the cotunneling current and TMR is displayed in Fig.
\ref{Fig:3}.

It is interesting to note that the angular dependence of electric
current and TMR differs now from that found above for the
situation (i). In order to understand this difference one should
take into account the fact that $\varphi_{\rm R}=\varphi_{\rm L}
=0$ corresponds to the parallel configuration, whereas the
situation with $\varphi_{\rm R}=-\varphi_{\rm L} =\pi /2$
corresponds to the antiparallel configuration with magnetic
moments oriented perpendicularly to the molecular field acting on
the dot. It is also worth noting that electric current reaches
minimum and TMR maximum not exactly at $\varphi_{\rm
R}=-\varphi_{\rm L} =\pi /2$, but for the configuration which is
close to the antiparallel one. Moreover, position of these minima
and maxima depends on the bias voltage, as shown in Fig. 3(c,d).
In turn, the case $\varphi_{\rm R} =-\varphi_{\rm L} =\pi$
corresponds again to the parallel configuration, but with the
magnetic moments of the electrodes being antiparallel to the
molecular field. This leads to a local maximum of electric current
(minimum in TMR) at $\varphi_{\rm R}=-\varphi_{\rm L} =\pi$. It is
interesting to note that the two parallel configurations (aligned
and anti-aligned with respect to the molecular field) are not
equivalent, and consequently the corresponding currents and also
TMR values are not equal.

When assuming equal spin polarizations of the leads and low bias
and temperature limits, one finds that the TMR effect at
$\varphi_{\rm R}=-\varphi_{\rm L} =\pi$ (which corresponds to its
local minimum) is given by
\begin{equation}
  {\rm TMR}^{({\rm min})}=\frac{4P(\varepsilon_\downarrow^2-
  \varepsilon_\uparrow^2)}
  {(1+P)^2\varepsilon_\uparrow^2+(1-P)^2\varepsilon_\downarrow^2}.
\end{equation}
In the same limit and for a nonzero spin polarization $P$, one can
show that the two maxima of TMR appear at $\varphi=\arccos\alpha$
and $\varphi =2\pi-\arccos\alpha$, with $\alpha=-\Delta/
P(\varepsilon_\uparrow+ \varepsilon_\downarrow)$. In the limit of
$P=0$ TMR vanishes by definition. In the case when $|\Delta|<
P(\varepsilon_\uparrow+\varepsilon_\downarrow)$, the maxima appear
approximately at $\varphi_{\rm R}=-\varphi_{\rm L} =\pi/2$ and
$\varphi_{\rm R}=-\varphi_{\rm L} =3\pi/2$. The corresponding TMR
value is then equal
\begin{equation}
{\rm TMR}^{({\rm max})}=\frac{P\left[
    (2+P)\varepsilon_\downarrow-(2-P)\varepsilon_\uparrow\right]
  (\varepsilon_\uparrow+\varepsilon_\downarrow)}
  {\varepsilon_\uparrow^2+\varepsilon_\downarrow^2-
  2P^2\varepsilon_\uparrow\varepsilon_\downarrow}.
\end{equation}

\section{Cotunneling through a singly occupied dot}

By applying an external gate voltage to the dot, one can tune
position of the level energy and this way also the dot occupation.
When $\varepsilon_\sigma $ is negative and $\varepsilon_\sigma+U$
positive, the dot is singly occupied at equilibrium for
$\Gamma,k_{\rm B}T \ll
|\varepsilon_\sigma|,|\varepsilon_\sigma+U|$, and the system is in
the Coulomb blockade regime. As before, we will consider the two
situations (i) and (ii) defined in the previous section.

\subsection{Theoretical description}

In the case studied in the preceding section the dot was empty,
and the second-order current was mediated only by elastic
cotunneling processes. When the dot is singly occupied, the
cotunneling current can also flow due to inelastic cotunneling, in
which the electrons tunneling to and off the dot have opposite
spin orientations (and consequently also different energies due to
the level spin-splitting). These inelastic cotunneling processes
determine the occupation numbers of the dot.

The probabilities $p_\uparrow$ and $p_\downarrow$, that the dot is
occupied either by a spin-up or spin-down electron, can be
calculated from the following stationary master equation:
\begin{eqnarray}
  0&=&\sum_{\alpha,\beta={\rm L,R}}\Big(-\gamma_{\alpha\beta ,\uparrow\Rightarrow
  \downarrow}\, p_{\uparrow} +
  \gamma_{\alpha\beta ,\downarrow\Rightarrow \uparrow}\,
  p_{\downarrow} \Big),
\end{eqnarray}
and taking into account also the normalization condition,
$p_{\uparrow}+p_{\downarrow}=1$. Here, $\gamma_{\alpha\beta
,\sigma\Rightarrow \bar{\sigma}}$ denotes the rate of inelastic
cotunneling  from lead $\alpha$ to lead $\beta$ with the
simultaneous change of the dot spin from $\sigma$ to
$\bar{\sigma}\equiv -\sigma$. The inelastic cotunneling processes
can take place through one of the two virtual states,
$\ket{\Phi_0}$ and $\ket{\Phi_{\rm d}}$, which correspond to the
empty and doubly occupied dot, respectively. The corresponding
energies are $\varepsilon_0=0$ for the empty state and
$\varepsilon_{\rm d}= \varepsilon_{\uparrow}+
\varepsilon_{\downarrow}+U$ for the doubly occupied state.

The rate of inelastic processes which transfer a spin-majority
electron from the left lead to the spin-down level of the dot and
a spin-up electron from the dot to the spin-majority electron band
in the right lead is given by
\begin{eqnarray}
  \gamma_{{\rm LR},\uparrow\Rightarrow\downarrow}^{+\Rightarrow+}&=&
  \frac{2\pi}{\hbar} \iint d\varepsilon_{\rm L} d\varepsilon_{\rm R}
  \rho_{\rm L+}\rho_{\rm R+}|T_{\rm L+}|^2|T_{\rm R+}|^2\nonumber\\
  && \times\sin^2\frac{\varphi_{\rm L}}{2}\cos^2\frac{\varphi_{\rm
  R}}{2}
  f(\varepsilon_{\rm L})[1-f(\varepsilon_{\rm R})]\nonumber\\
  &&\times \left( \frac{1}{ \varepsilon_{\rm L}+\mu_{\rm
  L}-\varepsilon_{\downarrow}-U}- \frac{1}{\varepsilon_{\rm
  R}+\mu_{\rm R}-\varepsilon_{\uparrow}} \right)^2
  \nonumber\\
  &&\times \delta(\varepsilon_{\rm L}+\mu_{\rm L}-\varepsilon_{\rm
  R}- \mu_{\rm R} +\varepsilon_{\uparrow}-\varepsilon_{\downarrow}).
\end{eqnarray}

The total rate of inelastic cotunneling from the left to right
leads, which changes the dot state from $\ket{\Phi_\uparrow}$ to
$\ket{\Phi_\downarrow}$, can be found by summing up over the
spin-majority and spin-minority electrons,
\begin{equation}
  \gamma_{\rm LR ,\uparrow\Rightarrow\downarrow}=
  \sum_{\sigma,\sigma^\prime=+,-} \gamma_{{\rm LR}
  ,\uparrow\Rightarrow\downarrow}^{\sigma\Rightarrow\sigma^\prime}.
\end{equation}
The corresponding analytical expression can be derived in a
similar way as in the case of elastic cotunneling described in the
previous section and takes the form
\begin{eqnarray}
    \lefteqn{\gamma _{\rm L R, \uparrow \Rightarrow \downarrow}=
    \frac{f_{\rm B} (\mu_{\rm R}-\mu_{\rm
    L}-\varepsilon_{\uparrow}+\varepsilon_{\downarrow})}{h}
    }\nonumber\\
&&\times
    \left(
    \Gamma_{\rm L}^{+} \Gamma_{\rm R}^{+} \sin^2\frac{\varphi_{\rm
    L}}{2} \cos^2\frac{\varphi_{\rm R}}{2}+ \Gamma_{\rm
    L}^{+}\Gamma_{\rm R}^{-} \sin^2\frac{\varphi_{\rm
    L}}{2}\sin^2\frac{\varphi_{\rm R}}{2}
    \right.\nonumber\\
    &&\left.+ \Gamma_{\rm L}^{-}\Gamma_{\rm
    R}^{+}\cos^2\frac{\varphi_{\rm L}}{2}\cos^2\frac{\varphi_{\rm
    R}}{2}+ \Gamma_{\rm L}^{-}\Gamma_{\rm
    R}^{-}\cos^2\frac{\varphi_{\rm L}}{2}\sin^2\frac{\varphi_{\rm
    R}}{2}
    \right)\nonumber\\
    &&\times\bigg\{B_2(\varepsilon_{\uparrow}-\mu_{\rm R})-
    B_2(\varepsilon_{\downarrow}-\mu_{\rm L}) \nonumber\\
    &&+B_2(\varepsilon_{\uparrow}+U-\mu_{\rm R})-
    B_2(\varepsilon_{\downarrow}+U-\mu_{\rm L})\nonumber\\
    &&+\frac{2}{U}\big[ B_1(\varepsilon_{\uparrow}-\mu_{\rm R})-
    B_1(\varepsilon_{\downarrow}-\mu_{\rm L}) \nonumber\\
    && - B_1(\varepsilon_{\uparrow}+U-\mu_{\rm R})+
    B_1(\varepsilon_{\downarrow}+U-\mu_{\rm L})\big] \bigg\}.
\end{eqnarray}
The cotunneling rate $\gamma _{\rm L R, \downarrow\Rightarrow
\uparrow }$ can be calculated in a similar way. In turn, the rate
of elastic cotunneling can be found as described in section III.

Having calculated the cotunneling rates and the occupation
probabilities, one can determine the current flowing from the left
to right leads. The contributions $I_{\rm inel}$ and  $I_{\rm el}$
due to inelastic and elastic cotunneling, respectively, are given
by
\begin{eqnarray}
  I_{\rm inel} &=&-e\sum_{\sigma=\uparrow,\downarrow} \left( \gamma
  _{\rm LR, \sigma \Rightarrow \bar{\sigma}}-\gamma _{\rm R
  L, \sigma
  \Rightarrow \bar{\sigma}}\right) p_{\sigma }, \\
  I_{\rm el}&=&-e\sum_{\sigma=\uparrow,\downarrow}
  \sum_{\sigma',\sigma''=+,-}\left(\gamma_{{\rm LR},\sigma
  }^{\sigma' \Rightarrow \sigma'' }-\gamma _{{\rm R L},
  \sigma}^{\sigma' \Rightarrow \sigma''
  }\right)p_{\sigma}.\hspace{0.4cm}
\end{eqnarray}
The total cotunneling current $I$ is then equal
\begin{equation}
  I=I_{\rm inel}+I_{\rm el}.
\end{equation}
The analytical expression for the current in the case of a singly
occupied dot is cumbersome and will not be presented here.

\subsection{Numerical results}

\begin{figure}[t]
\begin{center}
  \includegraphics[width=1\columnwidth]{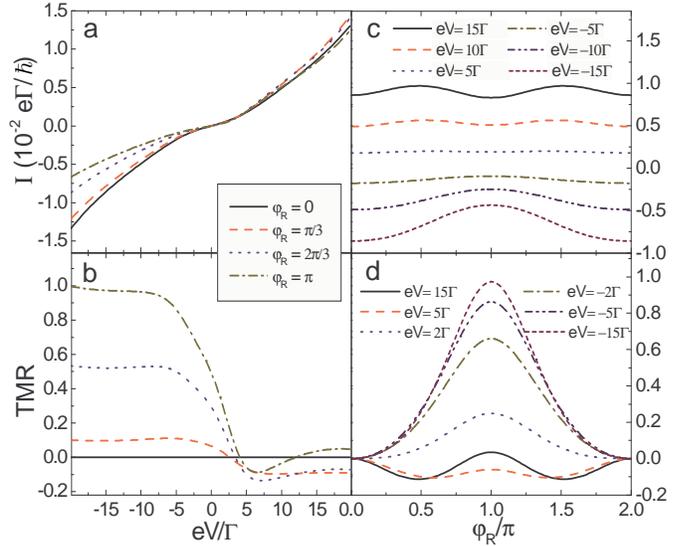}
  \caption{\label{Fig:4} (Color online) The cotunneling current
  (a,c) and the TMR effect (b,d) as a function of the bias voltage
  (left column) and the angle $\varphi_{\rm R}$ (the right column).
  The parameters assumed for calculations are: $k_{\rm
  B}T=\Gamma$, $\varepsilon_{\uparrow}=-22\Gamma$,
  $\varepsilon_{\downarrow}=-18\Gamma$, $U=40\Gamma$,
  $P_{\rm L}=P_{\rm R}=0.5$, $\varphi_{\rm L}=0$.}
\end{center}
\end{figure}

As far as physics is concerned, the situation with singly occupied
dot becomes more interesting. As before, we will analyze the two
situations (i) and (ii).

\subsubsection{Case (i): $\varphi_{\rm L}=0$}

When the magnetic moment of the left electrode is fixed
($\varphi_{\rm L}=0$) and the magnetic moment of the right lead is
free to rotate, the angular and bias dependence of electric
current and TMR is shown in Fig. 4. Except for the parallel
configuration, the current-voltage curves are now asymmetric with
respect to the bias reversal [see Fig. 4(a)]. This asymmetry also
leads to related asymmetric behavior of TMR [Fig. 4(b)]. Moreover,
for positive bias voltage, the TMR effect can change sign and
become negative in a certain range of the bias and angle values.
Such an asymmetry in transport characteristics with respect to the
bias reversal is of some importance for applications, particularly
when the current is significantly suppressed for one bias
polarization (diode behavior).

\begin{figure}[t]
\begin{center}
  \includegraphics[width=0.6\columnwidth,height=9cm]{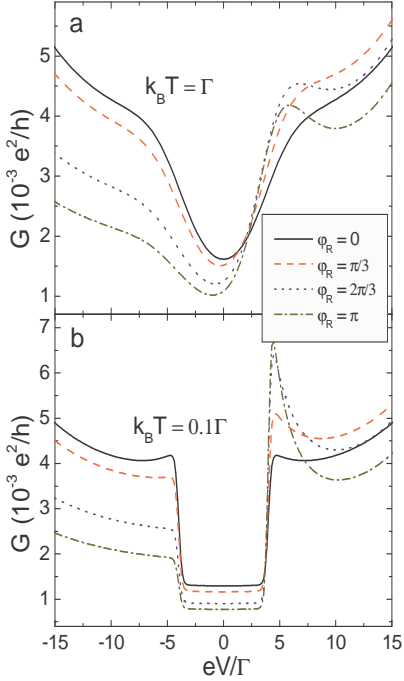}
  \caption{\label{Fig:5} (Color online) Differential conductance
  corresponding to the situation shown in Fig. 4(a), calculated for
  two different temperatures.}
\end{center}
\end{figure}

In order to account for the bias asymmetry, let us consider only
the antiparallel configuration. One should then realize that owing
to the level splitting, the single-barrier inelastic cotunneling
processes can occur only when the dot is occupied by a spin-down
electron. This follows simply from the energy conservation rule.
Thus, the single-barrier processes can assist the fastest
double-barrier cotunneling processes (spin-up electron tunnels
through the left barrier and spin-down electron tunnels through
the right barrier), but only for positive bias. This is because
the fastest processes can occur when the dot is occupied by a
spin-down electron for negative bias and by a spin-up electron for
positive bias. From this  follows that the conductance is larger
for positive than for negative bias voltage. This is indeed the
case in the characteristics shown in Fig. 4(a). The above
described mechanism of the asymmetry with respect to the bias
reversal does not hold when magnetic moments of the two leads are
parallel, so the corresponding current-voltage curve is
symmetrical.

To demonstrate the above described asymmetry more clearly, the
corresponding differential conductance is displayed in Fig. 5(a).
The asymmetry for $\varphi_{\rm R} > 0$ is clearly evident.
Besides the asymmetry, an additional interesting feature of the
differential conductance is also visible, namely the
characteristic deep in the small bias regime. This deep is a
consequence of the suppression of inelastic double-barrier
cotunneling events when $\vert eV\vert<|\Delta|$. For $\vert
eV\vert>|\Delta|$, the inelastic cotunneling processes are
allowed, leading to an enhanced conductance. The asymmetry and
zero bias anomaly are even more evident at lower temperature, as
shown in Fig. 5(b). Such a suppression of the inelastic
cotunneling events at small bias was used recently as a
spectroscopic tool to determine spin splitting of the dot level
and the corresponding $g$-factor \cite{kogan04}.

\begin{figure}[t]
\begin{center}
  \includegraphics[width=0.6\columnwidth,height=9cm]{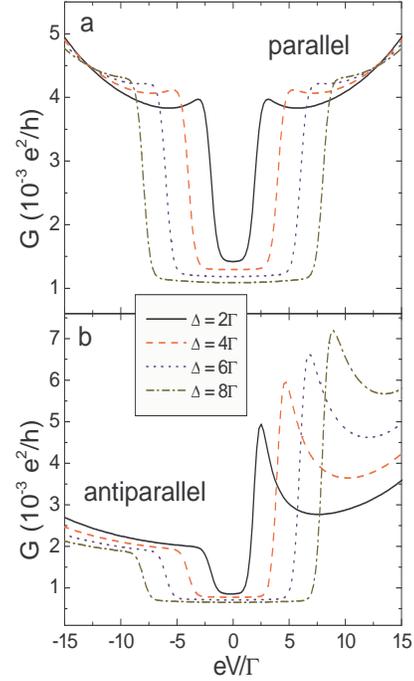}
  \caption{\label{Fig:6} (Color online) Differential conductance for
  different values of the level splitting for the parallel (a) and
  antiparallel (b) magnetic configurations. The parameters are:
  $k_{\rm B}T=0.2\Gamma$, $\varepsilon_\uparrow=\varepsilon-\Delta/2$,
  $\varepsilon_\downarrow=\varepsilon+\Delta/2$,
  $\varepsilon=-20\Gamma$, $U=40\Gamma$, and
  $P_{\rm L}=P_{\rm R}=0.5$.}
\end{center}
\end{figure}

The angular variation of electric current and TMR reveals further
new features. For negative bias there is a maximum of absolute
value of electric current in the parallel configuration and a
minimum in the antiparallel configuration. For positive bias,
however, the electric current has a maximum for noncollinear
configuration, as depicted in Fig. 4(c). The nonmonotonous
variation of electric current with the angle $\varphi_{\rm R}$
leads to the corresponding nonmonotonous variation of TMR, shown
in Fig. 4(d). Furthermore, TMR may now become negative, as already
mentioned before.

Numerical results presented so far were shown for a single value
of the level splitting, $\Delta=4\Gamma$. From the experimental
point of view, variation of the conductance as a function of the
level splitting (induced for instance by a strong external
magnetic field), allows one to determine some interesting
transport and spectroscopic characteristics. Therefore, in Fig. 6
we show the differential conductance for different splitting of
the dot level in the parallel and antiparallel configurations. By
measuring width of the conductance deep, one can determine for
instance the spectroscopic $g$-factor \cite{kogan04}.

In the case of a deep Coulomb blockade regime and $|eV|,k_{\rm B}T
\ll |\Delta|$, one can derive an approximate formula for the deep
in differential conductance due to the suppression of inelastic
cotunneling. Writing $\varepsilon_\uparrow=\varepsilon-\Delta/2$
and $\varepsilon_\downarrow=\varepsilon+\Delta/2$  one finds then
the following expression:
\begin{eqnarray}\label{Eq:GPlateau}
  G&=&\frac{\Gamma^2 e^2}{4h}\left[ \frac{(1+P_{\rm
  L}\cos\varphi_{\rm L})
  (1+P_{\rm R}\cos\varphi_{\rm R})}{(\varepsilon-\Delta/2)^2} \right.\nonumber\\
  &&+\frac{(1-P_{\rm L}\cos\varphi_{\rm L})
  (1-P_{\rm R}\cos\varphi_{\rm R})}{(\varepsilon+U+\Delta/2)^2}\nonumber\\
  &&\left.-\frac{8P_{\rm L}P_{\rm R}\sin\varphi_{\rm L}
  \sin\varphi_{\rm R}}
  {(\varepsilon-\Delta/2)(\varepsilon+U+\Delta/2)}\right],
\end{eqnarray}
which is valid for arbitrary magnetic configurations. This
expression approximates the plateaus shown in Fig. 5. When
assuming $\varphi_{\rm L}=\varphi_{\rm R}=0$ (which corresponds to
the parallel configuration), Eq. (\ref{Eq:GPlateau}) simplifies to
the following form
\begin{equation}
  G_{\rm P}=\frac{\Gamma^2 e^2}{4h}\left[ \frac{(1+P_{\rm
  L})(1+P_{\rm R})}{(\varepsilon-\Delta/2)^2}
  +\frac{(1-P_{\rm L})(1-P_{\rm
  R})}{(\varepsilon+U+\Delta/2)^2}\right],
\end{equation}
whereas for the antiparallel alignment ($\varphi_{\rm L}=0$,
$\varphi_{\rm R}=\pi$) it becomes
\begin{equation}
  G_{\rm AP}=\frac{\Gamma^2 e^2}{4h}\left[ \frac{(1+P_{\rm
  L})(1-P_{\rm R})}{(\varepsilon-\Delta/2)^2}
  +\frac{(1-P_{\rm L})(1+P_{\rm
  R})}{(\varepsilon+U+\Delta/2)^2}\right].
\end{equation}
The above two expressions describe the plateaus in differential
conductance shown in Fig. 6. It is also worth noting that
generally $G_{\rm P}>G_{\rm AP}$. Furthermore, both $G_{\rm P}$
and $G_{\rm AP}$ vary monotonously with spin polarization of the
leads -- in the case of nonmagnetic leads ($P_{\rm L}=P_{\rm
R}=0$) $G_{\rm P}=G_{\rm AP}$, whereas for $P_{\rm L}=P_{\rm R}=1$
(which corresponds to half-metallic leads) $G_{\rm P}$ is maximal
and $G_{\rm AP}=0$.

\subsubsection{Case (ii): $\varphi_{\rm R}=-\varphi_{\rm L}$}

\begin{figure}[b]
\begin{center}
  \includegraphics[width=1\columnwidth]{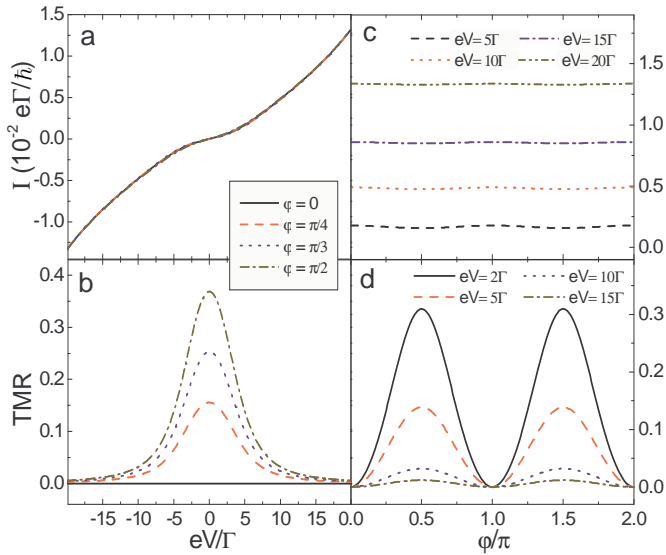}
  \caption{\label{Fig:7} (Color online) The cotunneling current
  (a,c) and the TMR effect (b,d) as a function of the bias voltage
  (left column) and $\varphi=\varphi_{\rm R}=-\varphi_{\rm L}$
  (right column). The other parameters are the same as in Fig. 4.}
\end{center}
\end{figure}

Transport characteristics in the second situation, i.e., for the
case when $\varphi_{\rm R}=-\varphi_{\rm L}=\varphi$, are
displayed in Fig. 7. One can note that the current is now almost
independent of the magnetic configuration. Nevertheless, the
angular dependence of the current becomes more visible in the
corresponding differential conductance, plotted in Fig. 8(a) for
different values of the angle $\varphi$, and also in Fig. 8(b) for
the same situation, but for much lower temperature. The
cotunneling gap due to suppression of the inelastic processes is
also clearly visible. The deep in differential conductance for
$|eV|\ll k_{\rm B}T,|\Delta|$ is given by Eq. (\ref{Eq:GPlateau}).
Since the system is now symmetric, the current-voltage curves (and
consequently also the differential conductance) are symmetric with
respect to the bias reversal.

\begin{figure}[t]
\begin{center}
  \includegraphics[width=0.6\columnwidth,height=9cm]{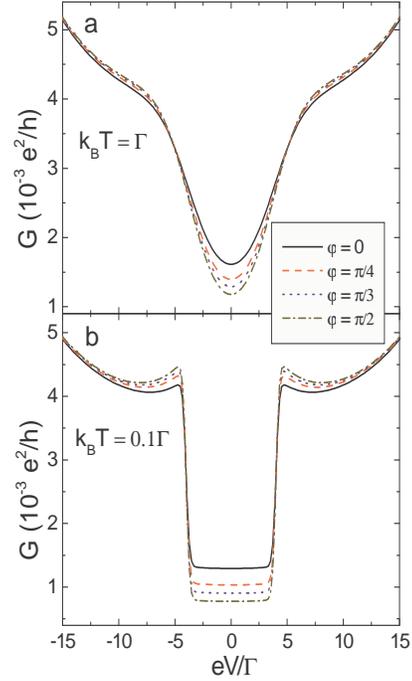}
  \caption{\label{Fig:8} (Color online) Differential conductance for
  the situation shown in Fig. 7(a), calculated for two indicated
  temperatures.}
\end{center}
\end{figure}

The TMR effect reaches maximum in the zero bias limit, $V=0$, as
shown in Fig. 7(b). In turn, angular variation of TMR reveals two
maxima (and also two minima), similarly as it was in the case of
empty dot, but now the maxima appear strictly for $\varphi=\pi/2$
and $\varphi=3\pi/2$. On the other hand, one minimum of TMR occurs
at $\varphi_{\rm R}=\varphi_{\rm L}=0$, where TMR vanishes by
definition. Tunnel magnetoresistance vanishes also in the second
parallel configuration, when both magnetizations are antiparallel
to the molecular field ($\varphi_{\rm R}=-\varphi_{\rm L}=\pi$).
This is due to the fact that the parameters assumed for numerical
calculations correspond to a symmetrical Anderson model, i.e.,
$U=-\varepsilon_\uparrow-\varepsilon_\downarrow$. When the system
becomes asymmetric, e.g., when $U$ increases (decreases) while the
other parameters are constant, the minimum in TMR at $\varphi
=\pi$ is shifted down (up) and when $U\gg
-\varepsilon_\uparrow-\varepsilon_\downarrow$, TMR has only one
maximum at $\varphi=\pi$, as shown in Fig. 9(b). On the other
hand, if $U< -\varepsilon_\uparrow- \varepsilon_\downarrow$, the
minimum in tunnel magnetoresistance at $\varphi=\pi$ becomes
negative, which is shown in Fig. 9(b) for $U=35\Gamma$. Thus, by
changing the model parameters one may significantly enhance or
reduce the TMR effect.

\begin{figure}[t]
\begin{center}
  \includegraphics[width=0.6\columnwidth,height=9cm]{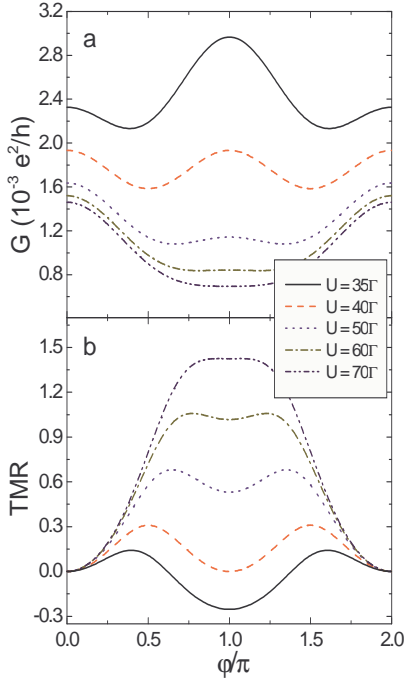}
  \caption{\label{Fig:9} (Color online) The angular dependence of
  the differential conductance and TMR for several values of the
  Coulomb interaction parameter $U$ at the bias voltage
  $eV=2\Gamma$. The other parameters are the same as in Fig. 7.}
\end{center}
\end{figure}

In the limit of $|eV|,k_{\rm B}T\ll|\Delta|$ and for $P_{\rm
L}=P_{\rm R}=P$, the TMR ratio for $\varphi_{\rm R}=-\varphi_{\rm
L}=\pi$ can be expressed as
\begin{equation}
  {\rm TMR}=\frac{4P(2\varepsilon+U)(U+\Delta)}
  {(1-P)^2(\varepsilon+U+\Delta/2)^2+(1+P)^2
  (\varepsilon-\Delta/2)^2} \, .
\end{equation}
From the above expression follows that the sign of TMR at
$\varphi=\pi$ depends on the ratio $\varepsilon/U$. If
$U=-2\varepsilon$, TMR vanishes, whereas for
$U\gtrless-2\varepsilon$, TMR is positive (negative). In the case
of $U\gg-2\varepsilon$, the TMR effect is given by $4P/(1-P)^2$.

The corresponding behavior of the differential conductance is
shown in Fig.9(a). Now, the maximum in the conductance at
$\varphi=\pi$ for a symmetric model changes into a minimum for
$U\gg -\varepsilon_\uparrow- \varepsilon_\downarrow$.

\section{Conclusions}

We have considered analytically and numerically cotunneling
current and associated tunnel magnetoresistance through a
single-level quantum dot coupled to two external ferromagnetic
leads. The dot level was assumed to be spin-split due to an
effective molecular field originating from a magnetic substrate on
which the dot is deposited, and the splitting was assumed to be
larger than the characteristic parameter $\Gamma$ (level width)
describing the dot-lead interaction. Although the general formulae
were derived for arbitrary in-plane orientation of the molecular
field and of the magnetic moments of the leads, detailed numerical
analysis was performed for two particular configurations, which
seem to be of particular interest. At this point we would like to
note, that a related problem has been recently studied by Pedersen
et al. \cite{flensberg04}, who considered cotunneling current in a
similar system, but for different magnetic geometry. In our case
all magnetic moments and molecular field were in plane of the
structure, whereas in their report magnetic moments were parallel
while magnetic field was tilted out of the plane.

We have found several interesting features in the angular and bias
dependence of transport characteristics. In the case of an empty
dot, TMR was found to be roughly independent of the bias voltage,
but strongly dependent on the angle between magnetic moments. When
magnetic moments of both leads rotate in opposite directions, both
electric current and TMR vary nonmonotonously with increasing
angle between the magnetic moments of the leads, and maximum of
TMR may occur at a noncollinear configuration. For a singly
occupied dot and for the case (i) ($\varphi_{\rm L}=0$), we found
strong asymmetry in electric current and TMR with respect to the
bias reversal, which disappears for the case (ii) ($\varphi_{\rm
R}=-\varphi_{\rm L}$). This diode-like behavior of the
current-voltage characteristics may be of some interest from the
application point of view. Moreover, the asymmetry in
current-voltage curves leads to associated asymmetry in TMR which
may become even negative for one bias polarization. An important
and interesting result is also an enhancement of TMR due to the
dot level splitting. Finally, we have also demonstrated
numerically and analytically the evolution of the cotunneling gap
with the splitting of the dot level and magnetic configuration of
the system.

The work was supported by the Polish State Committee for
Scientific Research through the projects PBZ/KBN/044/P03/2001 and
2 P03B 116 25. The authors acknowledge discussions with J\"urgen
K\"onig and Jan Martinek.

\appendix

\section{Calculation of cotunneling rate}

Here we present some details of the calculation of cotunneling
rates. As an example we consider the rate given by Eq. (4). Using
the properties of the delta-Dirac function it can be written as
\begin{eqnarray}
  \gamma_{\rm LR,0}^{+ \Rightarrow +}&=&
  \frac{\Gamma_{\rm L}^+ \Gamma_{\rm R}^+}{h}
  \int d\varepsilon f(\varepsilon)[1-f(\varepsilon+\mu_{\rm L}-\mu_{\rm R})] \nonumber\\
  &&\times\left[ \frac{\cos^2{(\varphi_{\rm L}/2)}\cos^2{(\varphi_{\rm
  R}/2)}} {(\varepsilon+\mu_{\rm L}-\varepsilon_{\uparrow})^2}\right.\nonumber\\
  &&+ \frac{\sin{\varphi_{\rm L}}\sin{\varphi_{\rm R}}}
  {2(\varepsilon_{\uparrow}-\varepsilon_{\downarrow})}\left(
  \frac{1}{\varepsilon+\mu_{\rm L}-\varepsilon_{\uparrow}}-
  \frac{1}{\varepsilon+\mu_{\rm L}-\varepsilon_{\downarrow}}
  \right)  \nonumber\\
  &&\left. +
  \frac{\sin^2{(\varphi_{\rm L}/2)}\sin^2{(\varphi_{\rm R}/2)}}
  {(\varepsilon+\mu_{\rm L}-\varepsilon_{\downarrow})^2}
  \right] \;.
\end{eqnarray}
Then, taking into account the identity
\begin{equation}
f(\varepsilon)[1-f(\varepsilon+ \xi^\prime)]=f_{\rm
B}(-\xi^\prime)[f(\varepsilon+\xi^\prime) -f(\varepsilon)]\;,
\end{equation}
Eq. (A1) can be rewritten as a sum of integrals of the type $\int
d\varepsilon f(\varepsilon)/ (\varepsilon-\xi)^n$ $(n=1,2,\dots)$.
In principal, it is necessary to determine the integrals for
$n=1$, whereas the ones for $n>1$ can be found using the
expression
\begin{equation}
\int d\varepsilon \frac{f(\varepsilon)} {(\varepsilon-\xi)^{n+1}}
= \frac{1}{n!} \frac{d^{(n)}}{d^{(n)}\xi}\int d\varepsilon
\frac{f(\varepsilon)} {\varepsilon-\xi} \;.
\end{equation}
Below we demonstrate the calculation of one of the summands,
namely
\begin{equation}
J=\int d\varepsilon \frac{f(\varepsilon)}{\varepsilon+\mu_{\rm L}-
\varepsilon_\uparrow} \;.
\end{equation}
To calculate this integral we use the Lorentzian cutoff of the
form $g_\alpha(\varepsilon)=W^2/ [(\varepsilon-\mu_\alpha)^2
+W^2]$, with $W$ being the cutoff parameter. Thus, Eq. (A3) can be
expressed as
\begin{eqnarray}
  J&=&\frac{W}{2i}\int d\varepsilon
  \frac{f(\varepsilon)}{(\varepsilon+\mu_{\rm L}-
  \varepsilon_\uparrow)(\varepsilon-\mu_{\rm L}-iW)}\nonumber\\
  &&- \frac{W}{2i}\int d\varepsilon
  \frac{f(\varepsilon)}{(\varepsilon+\mu_{\rm L}-
  \varepsilon_\uparrow)(\varepsilon-\mu_{\rm L}+iW)} \;.
\end{eqnarray}
The first (second) integral in the above formula has poles at
$\varepsilon=\varepsilon_\uparrow-\mu_{\rm L}$,
$\varepsilon=\mu_{\rm L}+(-)iW$, and $\varepsilon=i(2m+1)\pi$,
with $m=0,1,2,\dots$. However, because we are interested in the
deep Coulomb blockade regime (where the second-order processes
dominate), it is justifiable to assume $\varepsilon\ll
\varepsilon_\sigma$ and this way neglect the contribution of the
first pole. Then, by means of the contour integration and assuming
$W$ to be the largest energy scale, one gets
\begin{eqnarray}
  J&=&{\rm Re} \Psi{\left(\frac{1}{2}+
  i\frac{\varepsilon_\uparrow-\mu_{\rm L}}
  {2\pi k_{\rm B}T}\right)}-\ln{\left(\frac{W}{2\pi k_{\rm
  B}T}\right)}\;.
\end{eqnarray}
As the single integral depends on the cutoff parameter, the total
rate does not. The expressions depending on $W$ cancel in pairs,
which can be simply seen from Eq. (A2). The other summands of Eq.
(A1) can be found in a similar way with the aid of the above
mentioned identities.

Another way to calculate the cotunneling rates is to make use of
the assumption $\varepsilon\ll \varepsilon_\sigma$ and approximate
the resolvents of Eq. (A1) by $1/ (\varepsilon+ \mu_{\rm L}-
\varepsilon_\uparrow)\approx 1/ (\mu_{\rm L}-
\varepsilon_\uparrow)$. As a consequence, one arrives at the
expressions of type $\int d\varepsilon f(\varepsilon)
[1-f(\varepsilon+\xi^\prime)]$, which can be easily calculated
\cite{averin90}. The latter method is equivalent to the former one
if one expands the digamma functions and neglects the higher-order
corrections in $x/y$, with $x=|eV|,k_{\rm B}T$ and
$y=\varepsilon_\sigma,\varepsilon_\sigma+U$. The advantage of
using the first way of calculating the rates is that the
higher-order corrections in temperature are properly described.

\end{document}